\documentclass[aps,amsmath,amssymb,singlecolumn,nofootinbib,preprint,superscriptaddress]{revtex4-1}
\usepackage{graphicx}
\usepackage{bm}
\usepackage{lipsum}
\usepackage{epsfig}
\usepackage{epstopdf}
\usepackage{colordvi}
\usepackage{float}
\usepackage{accents}

\newcommand{\bean}{\begin{eqnarray*}}
\newcommand{\eean}{\end{eqnarray*}}
\newcommand{\ba}{\begin{array}}
\newcommand{\ea}{\end{array}}
\newcommand{\be}{\begin{equation}}
\newcommand{\ee}{\end{equation}}
\newcommand{\bea}{\begin{eqnarray}}
\newcommand{\eea}{\end{eqnarray}}

\newcommand{\no}{\nonumber}

\begin{document}
\title{Solitons supported by intensity-dependent dispersion}

\author{Chun-Yan Lin}
\affiliation{Institute of Photonics Technologies, National Tsing Hua University, Hsinchu 300, Taiwan}
\author{Jen-Hsu Chang}
\affiliation{Graduate School of National Defense, National Defense University, Taoyuan city 335, Taiwan}
\author{Gershon Kurizki}
\affiliation{Department of Chemical Physics, Weizmann Institute of Science, Rehovot 7610001, Israel}
\author{Ray-Kuang Lee}
\affiliation{Institute of Photonics Technologies, National Tsing Hua University, Hsinchu 300, Taiwan}
\affiliation{Physics Division, National Center of Theoretical Sciences, Hsinchu 300, Taiwan}

%

\begin{abstract}
Soliton solutions are studied for paraxial wave propagation with intensity-dependent dispersion. Although the corresponding Lagrangian density has a singularity, analytical solutions, derived by the pseudo-potential method and the corresponding phase diagram, exhibit one- and two-humped solitons with almost perfect agreement to numerical solutions. The results obtained in this work reveal a hitherto unexplored area of soliton physics associated with nonlinear corrections to wave dispersion.
\end{abstract}

\maketitle

Chromatic dispersion is the dependence of the phase velocity of a wave on its frequency~\cite{book} or, equivalently, frequency dependence of the refractive index. Nonlinear corrections to the chromatic dispersion as a function of the wave intensity arise for various waves, such as shallow water waves~\cite{Whitham, Whitham-book}, acoustic waves in micro-inhomogeneous media~\cite{acoustic}, or ultrafast coherent pulses in GaAs/AlGaAs quantum well waveguide structures~\cite{JMO}. In the context of photon-atom interactions, nonlinear dispersion effects may come about from the saturation of the atomic-level population~\cite{microwave}, electromagnetically-induced transparency (EIT) in a chain-$\Lambda$  configuration~\cite{EIT}, or nonlocal nonlinearity mediated by dipole-dipole interactions~\cite{sg}.

The interplay  between refractive-index nonlinearity and linear dispersion effects in a medium is expected to give rise to solitary, undistorted  wavepacket shapes over extended  travel  distance. However, soliton solutions of this kind are still unknown. Here, we search for soliton solutions in paraxial wave propagation along the axis $\zeta$, with an intensity-dependent dispersion:
\begin{eqnarray}
 i \frac{\partial \psi}{\partial \zeta} = \beta_2 (|\psi|^2) \, \frac{\partial^2 \psi(\zeta, \tau)}{\partial \tau^2},
 \end{eqnarray}
where $ \psi(\zeta, \tau)$ describes the envelope function of the wave, and $\beta_2 (|\psi|^2)$
denotes the intensity-dependent dispersion due to the interaction.

We may perform a Taylor expansion of the nonlinear dispersion term and restrict ourselves to the lowest-order quadratic correction whose strength is measured by the nonlinear coefficient $b$,  i.e.,
\begin{eqnarray}
 i \frac{\partial \psi}{\partial \zeta} =  \beta_2^0 (-1 + b\, |\psi|^2)\, \frac{\partial^2 \psi}{\partial \tau^2}.
 \end{eqnarray}
As $b = 0$, we have the wave propagating with the group velocity dispersion $\beta_2^0$, which is set to $1$ in the following.
The corresponding  Lagrangian density for Eq. (2) has the form  
\be \textit{L} =\frac{i (\bar{\psi_\zeta}\psi -\psi_\zeta \bar{\psi})} {2\,b\, \vert \psi \vert^2} \ln \vert -1+b\, \vert \psi \vert^2 \vert-\frac{1}{2} \vert \psi_\tau \vert^2. \label{lag}
\ee
We note that Eq. (2) also preserves the $\text{U}(1)$ symmetry, i.e., $ \psi \to \exp[i \theta] \psi$. 
From the Noether theorem~\cite{po}, one can obtain the conserved density for this model equation:
\be \frac{1}{b} \int_{-\infty}^{\infty} \ln \vert -1+b   \vert  \psi \vert^2 \vert d\tau.  \label{de} \ee 
For $b = 0$, the corresponding Lagrangian density given in Eq. (3), as well as the conserved density given in Eq. (4), both go to infinity. 
In this limit, we only have plane wave solutions supported by linear dispersion.

To find soliton solutions with a confined spatial profile, we adopt the stationary ansatz
\[
\psi=\text{X}(\tau)\,\exp[{i\, c\,\zeta}],
\]
with the real function $X(\tau)$ to be determined for a given propagation constant $c > 0$. By substituting this ansatz into Eq. (2), one has
\be -c\, \text{X}(\tau)=[-1+b\, \text{X}(\tau)^2]\, \text{X}^{\prime\prime}(\tau).\label{ja} \ee
By resorting to the concept of a pseudo-potentia~\cite{as, dr} i.e., $\text{X}^{\prime\prime} = - \nabla \text{V}(\text{X})$, we can find the corresponding pseudo-potential for the intensity-dependent dispersion in Eq. (2), to be
\begin{eqnarray}
\text{V}(\text{X})=\int \frac{c\, \text{X}} { -1+b\, \text{X}^2  } d \text{X}=\frac{c}{2b}\ln \vert-1+b\, \text{X}^2 \vert,
\end{eqnarray}
that vanishes at the origin, $\text{V}(\text{X} = 0)=0$. 
The potential in Eq. (6), must be a trapping potential in order to support bright solitons as bound states. That is, the pseudo-potential must  have either $b < 0$ to ensure  that it is  negative, or  $b  >  0$  and   $b\text{X}^2 < 2$.   

In the latter case,  the pseudo-potential  has  a  singularity  at  $\text{V}(\text{X})=0$,  for  X  >  0.   The amplitude of  the supported soliton is determined by $\text{V}(\text{X}) = 0$, so that  $\text{X} = 2/b$.
 For these two cases, we can obtain the solution $\text{X}(\tau)$ from Eq.(5) with the asymptotic condition $\text{X}(\tau \rightarrow \infty) = 0$,  by solving the Newtonian equation for a fictitious particle in the pseudo-potential, i.e., $\frac{1}{2}(\frac{d\text{X}}{d\tau})^2 = \text{V}(\text{X})$:
\begin{eqnarray}
&&\vert \tau-\tau_0  \vert =   \int_\text{X}^{M}  \sqrt{\frac{-b}{c} }  \frac{d \text{X}_1}{\sqrt{  \ln (1-b\, \text{X}_1^2) } }, \quad  b < 0;\\
&&\vert \tau-\tau_0  \vert =  \int_\text{X}^{M }  \frac{\sqrt{b}\, d \text{X}_1}{\sqrt{-c \ln \vert (-1+b\, \text{X}_1^2) \vert } }, \quad  b > 0.
\end{eqnarray}
Here, the maximum value  $\text{X}$ at $\tau_0 $ is assigned by $\text{X}(\tau_0) = M > 0$.
In both cases, when $\text{X} \approx 0$, one can also apply Taylor's expansion for  $\ln (1-b \text{X}^2) \approx -b \text{X}^2- b^2 \text{X}^4/2- b^3 \text{X}^6/3-\cdots $.
Then, as $\sqrt{\frac{\vert b \vert }{\vert c \vert } }  \frac{1}{\sqrt {  \ln \vert-1+b\, \text{X}^{2} \vert  } }  \approx \sqrt{\frac{1} {\vert c \vert }} \frac{1} {\text{X}}$,  we have $\text{X} \to  0$ as $ \tau \to \pm \infty$.
Due to the translation invariance,  we can set $\tau_0 = 0$ for $\text{X}(0) = M$.  Then, the corresponding derivative $\text{X}^\prime(0)$  can be obtained as
$\text{X}^\prime(0)= - \sqrt{\frac{c}{b} \ln \vert -1+b\, M^2 \vert}$. Now, we can find soliton solutions for Eq. (7) or (8) with $b <  0$ or $b > 0$, respectively.

\begin{figure}[t]
\includegraphics[width=8.0cm]{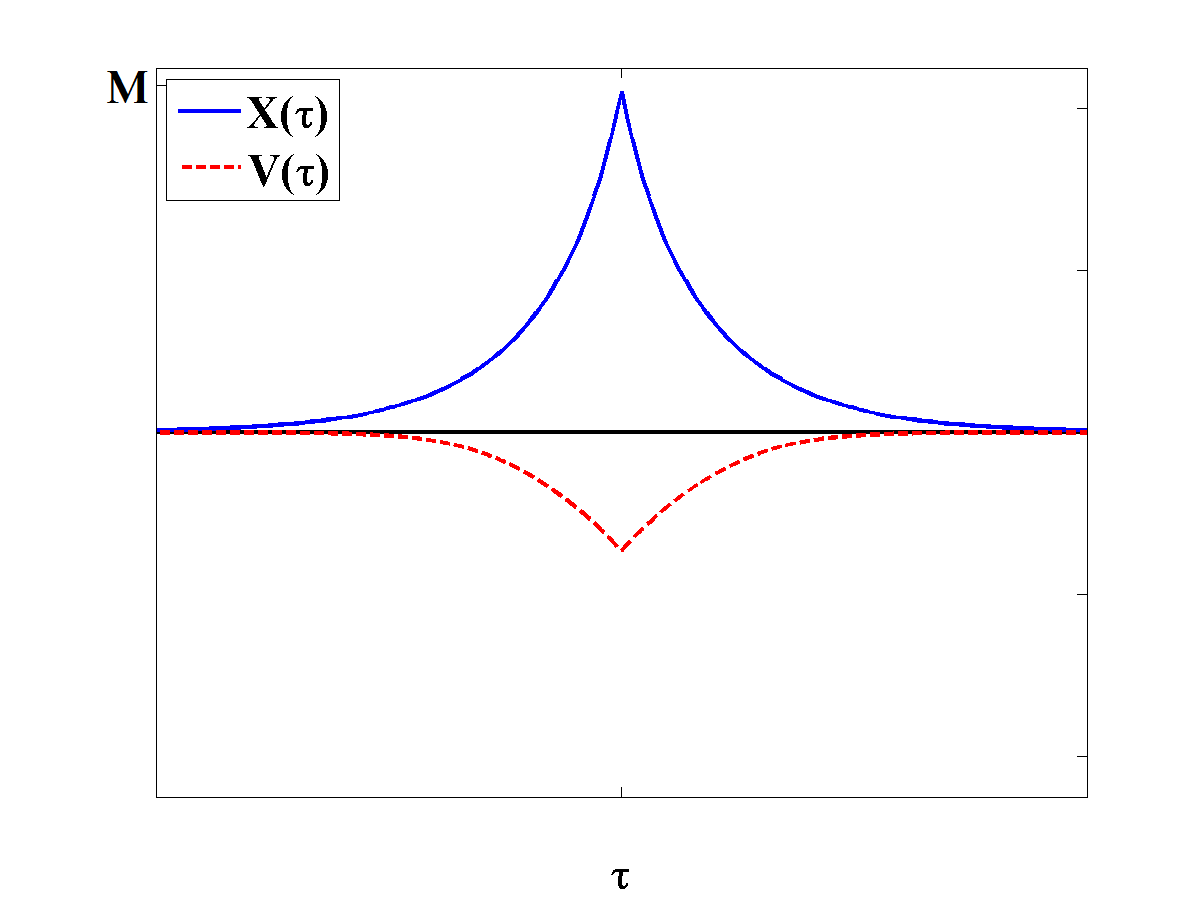}
\caption{The illustration of soliton solution $\text{X}(\tau)$ and its corresponding  pseudo-potential $\text{V}(\tau)$, depicted in solid- and dashed-curves, respectively.
 The soliton is supported by intensity-dependent dispersion, with a negative value of the nonlinear coefficient, $b = -1 < 0$. The soliton solution $\text{X}(\tau)$ derived in Eq.
(10) almost  perfectly  matches the numerical one obtained by directly solving  Eq. (2). Here, M denotes the maximum value of the soliton profile at $\tau = 0$, i.e., $\text{X}(\tau = 0) = M$, and $c = 1$.}	
\end{figure}

For a negative nonlinear coefficient, $b < 0$, one can match the asymptotics at $\text{X} \to \pm \infty $  with Taylor’s expansion near $\text{X} = 0$ and arrive at
\be 
\vert \tau \vert = -\ln  \left( \text{X} \right) {\frac {1}{\sqrt {c}}}+\frac{1}{8}\,{\frac {b}{\sqrt {c}}}{\text{X}}^{2}+ O(\text{X}^4).
\ee
Then, we have the following approximation for the corresponding soliton solution:
\be \text{X}(\tau) \approx \exp[{\frac{-1}{2}\,{\it W} \left( \frac{-1}{4}\,b\,{{\rm e}^{-2\, \sqrt {c}  \vert \tau \vert }} \right) -\sqrt {c}  \vert \tau \vert }],
\ee
where ${\it W}$ denotes the Lambert function defined as
\[ 
{\it W}(z)\, e^{{\it W}(z)} = z. 
\]
Equation (10) is the main result of this work: it yields the soliton profile supported only by intensity-dependent dispersion. One can see that  $\text{X}(\tau) \approx \exp[ - \sqrt {-c} \vert \tau \vert ]$ as $\vert \tau \vert \to \infty $ since ${\it W}(0) = 0$. 
It corresponds to the reduced linear equation in Eq. (2),  i.e., $b=0$. 
Moreover, as $ \vert \tau \vert \to 0  $, we have $\text{X}(0) = M = \exp[{\frac{-1}{2}\,{\it W} ( \frac{-1}{4}b )}]$. 
Then, if $ b \to -\infty $, we have $\text{X}(0) \to \infty $ as $\lim_{z \to \infty}{\it W}(z) \to \infty$.

A comparison between our analytical solution in Eq. (10) and a numerical solution obtained by directly solving Eq. (2) is depicted in Fig. 1. As shown by the solid curves, the soliton solution $\text{X}(\tau)$ derived in Eq. (10) almost perfectly matches the numerical ones obtained by directly solving Eq. (2). 
We also depict the corresponding pseudo-potential (dashed-curve) as $\text{V}(\tau)$, by setting $b = 1$. Due to the reflection symmetry, the function $\text{X}(\tau)$ for $\tau < 0
$ is constructed from Eq. (10) by taking $\text{X}( -\tau) = \text{X}(\tau)$. The maximum value of the soliton profile at $\tau = 0$ is set to be $\text{X}(\tau = 0) = M$. 
One can see that the derivative of the supported soliton profile diverges at the center, i.e., $d\text{X}(\tau= 0 )/d\tau = \pm \infty$. With the introduction of a non-zero nonlinear coefficient, the resulting pseudo-potential $\text{V}(\tau)$ acquires a discontinuity in its first-order derivative at $\tau = 0$.

\begin{figure}[t]
\includegraphics[width=0.45\textwidth]{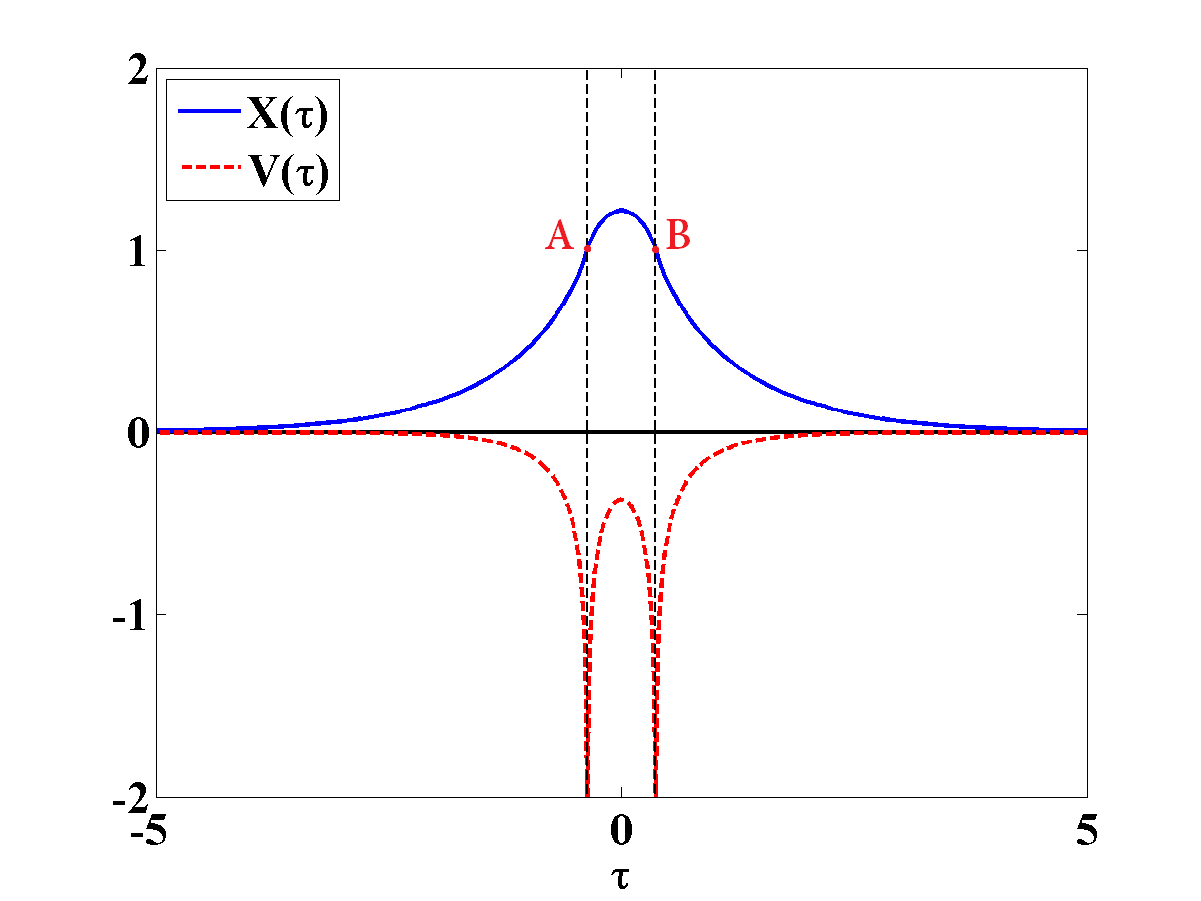}
\caption{
One-humped soliton solution $\text{X}(\tau)$ and the corresponding pseudo-potential $\text{V}(\tau)$, depicted by solid and dashed curves, respectively, for a positive value in the nonlinear coefficient, $b = +1 > 0$. The tails of the soliton solution $\text{X}(\tau)$ obtained numerically from Eq. (2) can be reconstructed  from Eq. (10). The derivatives of the supported soliton profile diverge at the two points marked A and B, where $d\text{X}/d\tau = \pm \infty$, and $c = 1$.}
\end{figure}

The Lambert function W(z) has the domain $z \in [-1/e, \infty)$, with the minimal value $-1$ at $z=- 1/e$. 
Hence,  in Eq. (10), we have $ -b \exp[{-2 \sqrt{c}\, \tau}]/4  \geq  - 1/ e$, or 
\be \tau   \geq \frac{ \vert \ln \vert \frac{b\,e}{4} \vert \vert }{ 2 \sqrt {c}} \approx \frac{ \ln (b)-0.3862}{ 2 \sqrt {c}}.
\ee
This result approximates the soliton solution given in Eq. (10), as $  \vert \tau \vert  \to  \infty$. 
In addition, when $-b \exp[{-2 \sqrt{c} x}]/4 = - 1/e$,  one has  the value 
\be 
\text{X}(\tau_0) = \frac{ \vert \ln \vert \frac{b\,e}{4} \vert \vert }{ 2 \sqrt {c}}) \approx  \sqrt{\frac{2}{b}},
\ee
 which is the maximum value for the amplitude of soliton solutions at $\tau_0$.

Based on above argument, we can set $M = \sqrt{2/b} >0$ for a positive nonlinear coefficient, i.e., $b > 0$.
In Fig. 2, we depict the numerical solutions for $\text{X}(\tau)$  by the solid curve, which is obtained by directly solving Eq. (2) with a positive value in the nonlinear coefficient, b = +1. Here, even- symmetry soliton solutions are constructed, i.e., $\text{X}(-\tau)=\text{X}(\tau)$. 
Except for the profile between  the two points marked A and B, the tails of the soliton solution $\text{X}(\tau)$ can be almost exactly reconstructed from Eq. (10). As the corresponding pseudo-potential $\text{V}(\tau)$ goes to $\infty$ at points A and B (see the dashed-curve), the derivatives of the supported soliton profile also diverge at these two points.

\begin{figure}[t]
\includegraphics[width=0.45\textwidth]{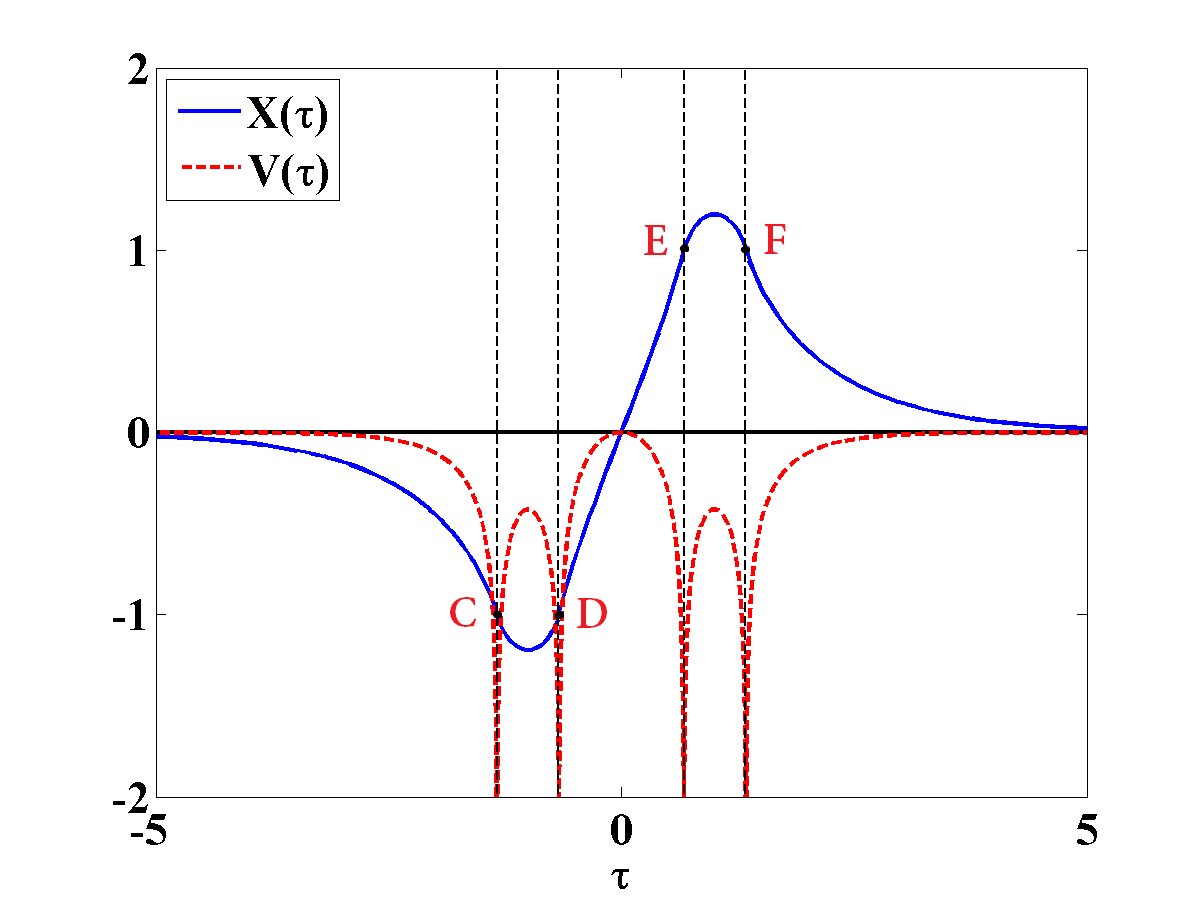}
\caption{Two-humped soliton solution $\text{X}(\tau)$ and its corresponding pseudo-potential $\text{V}(\tau)$, depicted by solid and dashed curves, respectively, with a positive value in the nonlinear coefficient, $b = +1 > 0$. Here, the derivatives of the supported soliton profiles diverge at the points marked C, D, E, and F, where $d\text{X}/d\tau = \pm \infty$.}
\end{figure}

Even though the supported soliton solution shown in Fig. 2 has points with divergent derivatives, one can prove that the corresponding conserved density still remains finite and thus the solution is physical. By using the relation between $\tau$ and $\text{X}$ given in Eq. (8), one can change the integral variable in Eq. (4)
\bea  
&&  \frac{1}{\sqrt{bc}} \int_{0}^{M} \sqrt {\ln   \vert -1+b\text{X}_1^2 \vert  }\, d \text{X}_1, \\\nonumber
&&= \frac{1}{\sqrt{ bc}}\left[ \int_{0}^{\frac{1}{\sqrt{b}}} \sqrt {\ln   (1-b\text{X}_1^2)  } d \text{X}_1 + \int_{ \frac{1}{\sqrt{b}}}^{M} \sqrt {\ln   (-1+b\text{X}_1^2)   } d\text{X}_1\right], \\\no 
&&=  \frac{1}{2b} [\int_{0}^{1} \frac{\sqrt{\frac{\ln u }{c}}}{\sqrt{1-u}} du  + \int_{0}^{bM^2-1}  \frac{\sqrt{\frac{\ln u_1 }{c}}}{\sqrt{1+ u_1}} d u_1  ],  \label{qq} 
\eea
where we have introduced $u\equiv 1-b \text{X}_1^2$ and $u_1\equiv -1+b \text{X}_1^2$. 
As it is known that $\lim_{u \to 1^{-}} \frac{\ln u}{1-u}=-1 $ by the $L'hopital$ rule, the convergence of improper integrals in Eq.  (13)  depends on the integral $\int_{0}^{\alpha} \sqrt{\frac{\ln v }{c}} dv$ for a finite positive $\alpha$ near $v=0$. 
By choosing $\alpha=1$ for the scaling, we have  
\[ \int_{0}^{1} \sqrt {  \frac{  \ln v }{c} } dv = \sqrt{ \frac{-1}{c}} \int_{0}^{\infty} y^{1/2} e^{-y} dy = \sqrt{ \frac{-1}{c}} \Gamma (3/2)=\sqrt{ \frac{-\pi}{4c}}, \]
with  $v=e^{-y}$. 
Hence, the conserved density of our stationary soliton solution is convergent even for a nonlinear dispersion coefficient $b > 0$.

In addition to the one-humped even-symmetry soliton solutions displayed in Fig. 2, we can also construct odd-symmetry two-humped soliton solutions for a positive value of the nonlinear coefficient, $b = +1$. One can see in Fig.3 the odd-symmetry soliton solution $\text{X}(\tau)$ depicted by a solid curve, i.e., $\text{X}(-\tau) = -\text{X}(\tau)$, upon setting $M = 0$. The corresponding potential $\text{V}(\tau)$, depicted by a dashed-curve, has four singular values at the points marked C, D, E, and F. We can check from Eq. (2) that a finite value of the conserved density exists for the two-humped soliton solution

\begin{figure}[t]
\includegraphics[width=0.45\textwidth]{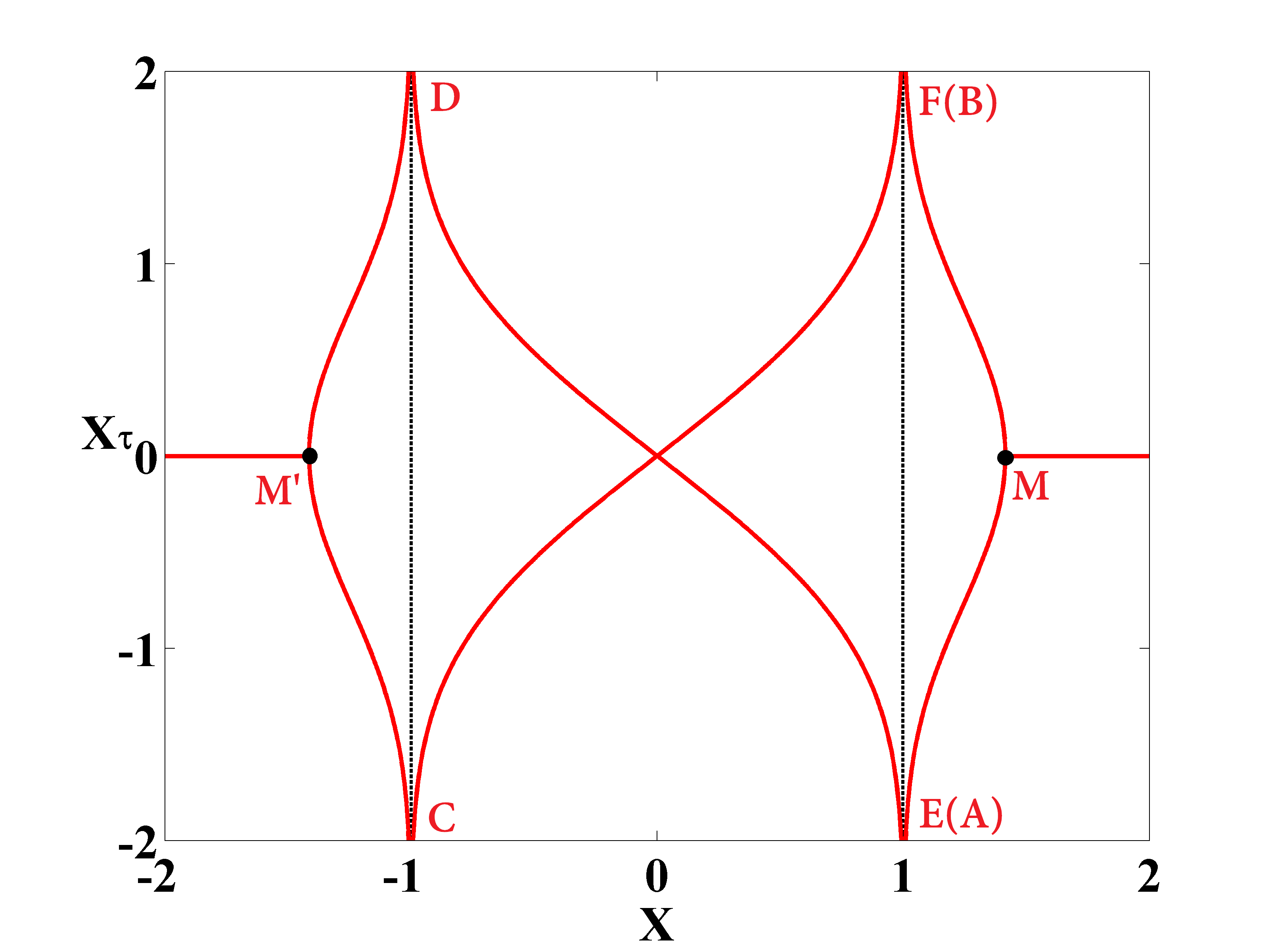}
\caption{The phase diagram defined by $\text{X}$ and $\text{X}_\tau \equiv d\text{X}/d \tau$ for our soliton solutions. Here, the sets of points marked (A, B) and (C, D, E, F) correspond to the marked points in Fig. 2, and 3, respectively; while the points M and $\text{M}^\prime$ give the maximum value of  the soliton profile. As $b = + 1$  is chosen, we have $\text{X} = \pm 1/\sqrt{b} = \pm 1$ for the amplitudes in the soliton profile, at which the derivative goes to $\pm \infty$.}
\end{figure}

An alternative picture that provides deeper understanding of our soliton solutions is obtained from the phase diagram for the Newtonian pseudo-particle dynamics, defined by $\text{X}$ and $\text{X}_\tau \equiv d\text{X}/d \tau$. 
For the one-humped solution, one may follow the trajectory on the right-hand side of this phase diagram, where $\text{X} \ge 0$ ( Fig. 4). 
By starting at the origin $(\text{X}, \text{X}_\tau) = (0, 0)$ and following the trajectory to the point marked B $(1/\sqrt{b}= 1, \infty)$, we find an infinite derivative of the profile.
The soliton profile goes through its maximum value (the point marked M) to its other infinite derivative (point marked A), and finally back to the origin $(0, 0)$. This trajectory exactly reflects the one-humped soliton solution illustrated in Fig. 2. By following the two sides of the trajectory in Fig. 4, $\text{X} \ge 0$ and $\text{X} \le 0$, one can easily construct the two-humped soliton solutions illustrated in Fig. 3.

In conclusion, our analysis reveals the existence of singularities in the pseudo-potential associated with intensity-dependent dispersion, resulting in one- and two-humped supported solitons with infinite derivatives in their profiles. The tails of these solitons can be described by Lambert functions, which give almost perfect agreement to the numerical solutions  of  the paraxial wave equation with intensity-dependent dispersion. Even though such discontinuities in the derivative of soliton profiles make them unstable (as we have checked by linear stability analysis and by the Vakhitov-Kolokolov stability criterion), their conserved density still remain finite, attesting to the physicality of the solutions. As nonlinear corrections to the dispersion arise in a variety of wave phenomena, our results may open a hitherto unexplored area of nonlinear wave propagation.
Through the correspondence between the paraxial wave equation and the Schrödinger equation (upon replacing  $\zeta \rightarrow t$ and  $\tau \rightarrow x$),  our model equation can also be applied to a quantum particle (electron or hole) with a nonlinear effective mass $m^\ast(|\psi|^2)$, i.e., $i\hbar \psi_t =    [1/2m^\ast(|\psi|^2)]\psi_{xx}$. In a nonuniform potential, a quantum particle may acquire a position-dependent effective mass. Such a scenario has gained much interest in view of its applications, ranging from semiconductors to quantum fluids~\cite{PDEM-1, PDEM-2, PDEM-3, PDEM-4, PDEM-5, AIP, cr}.

A number of promising applications and directions for further exploration may be identified: 
(a) The present soliton model may be connected to off-resonant electromagnetic (EM) propagation  in two-level media [6] outside the domain of resonant self-induced transparency (SIT) solitons~\cite{SIT, SIT2}. 
(b) In media with spatially-periodic refractivity doped with two-level systems (TLS) the spatial modulation of the propagating EM intensity may enhance the intensity-dependent nonlinear TLS dispersion ~\cite{TLA, TLA2}. 
(c) In the EIT regime of three-level atoms that are coupled via resonant dipole-dipole interactions, the present soliton solutions may be related to the previously explored long-range photon-photon interactions~\cite{3level, 3level2}.

This work is supported by the Ministry of Science and Technology of Taiwan under Grant No.: 105- 2628-M-007-003-MY4, 107-2115-M-606-001, and 108-2923-M- 007-001-MY3. G.K. acknowledges the support of EU FET Open (PATHOS), ISF, DFG (FOR 7024) and QUANTERA (PACE-IN).


\end{document}